\documentclass[reprint,preprintnumbers,superscriptaddress,amsmath,amssymb,floatfix,prl]{revtex4-1}

\usepackage{graphicx}           

\def\nm{{\ {\rm nm}}}                       

\def\THz{{\ {\rm THz}}}                     

\def\Er{{{E_r}}}                            
\def\kr{{{k_r}}}                            
\def\Rb87{^{87}\rm{Rb}}                     
\def\Li6{^{6}\rm{Li}}                       

\newcommand{\ud}{\mathrm{d}}

\begin{document}

\title{Spin-charge-density wave in a squircle-like Fermi surface for ultracold atoms}

\author{D.~Makogon}
\affiliation{Institute for Theoretical Physics, Utrecht University, 3508 TD Utrecht, The Netherlands}

\author{I.~B.~Spielman}
\affiliation{Joint Quantum Institute, National Institute of Standards and Technology, and University of Maryland, Gaithersburg, Maryland, 20899, USA}

\author{C.~Morais~Smith}
\affiliation{Institute for Theoretical Physics, Utrecht University, 3508 TD Utrecht, The Netherlands}


\date{\today}

\begin{abstract}
We derive and discuss an experimentally realistic model describing
ultracold atoms in an optical lattice including a commensurate,
but staggered, Zeeman field.  The resulting band structure is
quite exotic; fermions in the third band have an unusual rounded
picture-frame Fermi surface (essentially two concentric
squircles), leading to imperfect nesting.  We develop a
generalized ${\rm SO}(3,1)\times {\rm SO}(3,1)$ theory describing
the spin and charge degrees of freedom simultaneously, and show
that the system can develop a coupled spin-charge-density wave
order. This ordering is absent in studies of the Hubbard model
that treat spin and charge density separately.
\end{abstract}

\maketitle

{\it Introduction}
Ultracold atoms in optical lattices have recently emerged as a class of condensed matter systems, where the properties of the many-body
Hamiltonian are under exquisite experimental control.  Interfering laser beams in one, two or three dimensions (D) create standing waves:
nearly perfect optical lattices for atoms with lattice spacing and topology set by the laser geometry and wavelength~\cite{Bloch}.
Optical lattices not only allow for the implementation of different lattice models without defects, but also open a wide range of possibilities
to manipulate the parameters of the model describing ultracold bosons, fermions, or mixtures thereof.  For example, the hopping parameters,
local chemical potential, and often even the interaction strength can be tuned at will.


Most optical lattice experiments use atoms in a single
state~\cite{Greiner2002}, however, some experiments study mixtures
of atoms in two or more atomic ``spin'' states, each of which can
experience different lattice
potentials~\cite{Mandel2003,Lee2007,Lundblad2008}.  We derive a
lattice model, equally applicable to bosons and fermions, with an
effective Zeeman magnetic field including a term alternating in
sign on a site-by-site basis~\cite{Dudarev2004}. In condensed matter
systems, the Zeeman field
couples strongly to electrons near the Fermi
surface~\cite{Revaz_Ramazashvili}, and in more orchidaceous
situations, it breaks local time-reversal invariance in
topological insulators~\cite{Essin2009,Li2010}.

\begin{figure}[tb]
\begin{center}
\includegraphics[width=3.3in]{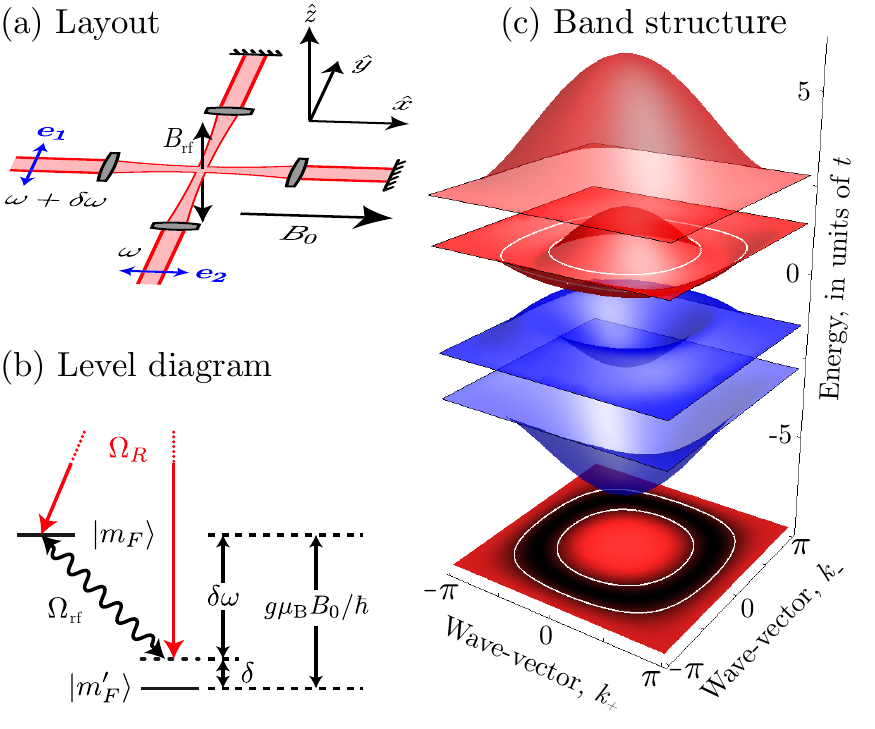}
\end{center}
\caption[Geometry and level diagram]{ (a) Schematic layout: two
nearly degenerate counter-propagating lasers differ in frequency
by a very small $\delta\omega$, and are linearly polarized in the
$\hat x$-$\hat y$ plane.  An rf magnetic field $B_{\rm rf}
\cos(\delta\omega t + \phi)$ is polarized along $\hat z$.  A bias
field $B_0 \hat x$ brings the two Zeeman-resolved $m_F$ levels
nearly into resonance. (b) Illustrative level diagram: the two
$m_F$ states are coupled by optical and rf magnetic fields.  The
lattice potential formed by the retro-reflected lasers is omitted.
(c) Band structure computed for $\Omega_{\rm R} = 2t$,
$\Omega_{\rm rf}=4t$, and $\phi=\pi/4$: the concentric-squircles
of the Fermi-surface are obtained by filling the system to the
third band, and are depicted by the white contours at the Fermi
energy (also shown is the projection onto the $k_+$-$k_-$
plane).}\label{geometry}
\end{figure}

For particles with two spin states, our lattice model has four
low-energy bands, and the third is shaped as a squarish, deformed,
Mexican hat for a wide range of system parameters.  By filling the
system with fermions, we obtain a peculiar Fermi surface,
consisting of the boundaries of a squarish ring, essentially two
concentric squircles~\cite{Guastia2005}.  The particular shape of
the Fermi surface suggests that nesting effects should be
expected.  To account for interactions, we develop an ${\rm
SO}(3,1)\times {\rm SO}(3,1)$ description of the charge and spin
degrees of freedom.  Imperfect nesting along the diagonal
connecting corners of the Fermi-surface gives rise to a coupled
spin-charge-density wave (SCDW) instability at a critical
interaction strength $U_c$.  We calculate the imaginary part of
the trace of the random phase approximation (RPA) susceptibility
to study the collective excitations of the system.  At the
interaction strength $U_c$, a soft mode arises at the optimal
nesting wave-vector ${\bf Q}$.  The SCDW instability is in general
incommensurate with the lattice, and is tunable by external
parameters.  In contrast with the usual behavior in 1D, our
results show that in 2D a combined treatment of spin and charge
degrees of freedom is essential to capture the possible
instabilities of the system.

The physical system under study [Figs.~\ref{geometry}(a)-(b)]
consists of a sample of ultra cold atoms illuminated by two pairs
of counter-propagating lasers with angular frequencies $\omega$
and $\omega+\delta\omega$ (where $\omega\gg\delta\omega$); a third
pair of lasers, not shown, propagate along $\pm\hat z$ and create
a 1D lattice, confining motion to the $\hat x-\hat y$ plane.  Our
model includes a static magnetic field $B_0$ along $\hat x$ and an
rf magnetic field $B_{\rm rf}$ with angular frequency $\delta
\omega$ along $\hat z$; rf coupling between atoms in spin
dependent lattices has been studied both
experimentally~\cite{Lundblad2008} and
theoretically~\cite{Yi2008}, where the resulting non-trivial
real-space lattices suggested potential application to many body
systems and quantum computation.  In our case, the spin dependence
results from the interplay of the laser and rf-magnetic fields.

As was observed in
Refs.~\cite{Deutsch1998,Dudarev2004,Sebby-Strabley2006},
conventional spin independent (scalar) optical lattice potentials
acquire additional spin-dependent terms near atomic resonance: the
rank-1 and rank-2 tensor light shifts~\cite{Deutsch1998}.  In the
case of alkali atoms, adiabatic elimination of the angular
momentum ${\mathcal J} =1/2$ (D1) and ${\mathcal J}=3/2$ (D2)
excited states yields an effective Hamiltonian $H_0=U_s({\bf
e}^*\cdot{\bf e}) + i U_v {\vec {\mathcal J}} \cdot({\bf e}^*\times{\bf
e})/\hbar$ for the ${\mathcal J}=1/2$ ground state atoms.  $\bf e$
is the polarization vector of the optical electric field and the
magnitude of the scalar and vector light shifts are related by
$U_v=-2U_s\Delta_{\rm FS}/3(\omega-\omega_0)$.  Here, the
fine-structure splitting is $\Delta_{\rm FS} =
\omega_{3/2}-\omega_{1/2}$; $\hbar\omega_{1/2}$ and
$\hbar\omega_{3/2}$ are the D1 and D2 transition energies; and
$\omega_0=(2\omega_{1/2}+\omega_{3/2})/3$ is their suitable
average.  $U_v$ and $U_s$ can be independently specified with
informed choices of laser frequency $\omega$ and intensity.  We
focus on a practical case, where the lasers are detuned far below
atomic resonance $\omega_0-\omega \gg \Delta_{\rm FS}$, minimizing
spontaneous emission and implying $|U_s|\gg|U_v|$ and $U_s < 0$.
We express momentum and energy in dimensions of
$\hbar\kr=\hbar\omega/c$ and $\Er=\hbar^2\kr^2/2 m$, the
single-photon recoil momentum and energy, respectively, with $m$
the atomic mass. 

The atomic Hamiltonian for the laser and magnetic fields in
Figs.~\ref{geometry}(a)-(b) is $H_0=U_s(\cos^2{\kr x} + \cos^2{\kr
y}) +g\mu_B  {\vec {\mathcal J}}\cdot {\bf B}_{\rm eff}$ with ${\bf
B}_{\rm eff} = B_{0}\hat x + B_{\rm rf}\cos(\delta\omega t + \phi)
\hat z + \hat z (U_v/2 g \mu_B) \cos(\delta\omega t)\cos(\kr
x)\cos(\kr y)$, where the vector light shift acts as an effective
magnetic field and $B_{0}\gg B_{\rm rf},U_v/2 g \mu_B$.  Here,
$\mu_{\rm B}$ is the Bohr magneton and $g$ is the Land\'e
$g$-factor.  We select $\hat x$ as the quantizing axes, transform
into the frame rotating at $\delta \omega$, and make the rotating
wave approximation to find
\begin{align*}
{\bf B}_{\rm eff} &= \left(B_{0}-\frac{\hbar\delta\omega}{g\mu_B}\right) \hat x - \frac{B_{\rm rf}}{2}\sin(\phi) \hat y + \\
& \left[\frac{B_{\rm rf}}{2}\cos(\phi) + \frac{U_v}{4 g \mu_B}\cos(\kr x)\cos(\kr y)\right]\hat z.
\end{align*}
${\bf B}_{\rm eff}\cdot\hat z$ reaches its extrema on the sites of
the optical lattice, giving a bias plus staggered Zeeman field.
This proposal requires the simple retro-reflection of the existing
``Raman'' lasers discussed in Ref.~\cite{Lin2009b}, which were
used to create an artificial magnetic field (there, $B_{\rm rf}$
was used only for state preparation).  When $\left|U_s\right|\gg
\left|U_v\right|, \left|g\mu_B B_{\rm rf}\right|$ the conventional
tight-binding model~\cite{Jaksch1998}, valid when
$U_s\gtrsim5\Er$, is slightly modified by the effective magnetic
field evaluated on the lattice sites, yielding
\begin{align*}
H_0 &= -t\sum_{\left<{\bf i},{\bf j}\right>,s}c^\dagger_{{\bf i},s} c_{{\bf j},s}+\frac{\Omega_{\rm rf}}
{2}\sum_{{\bf j}}\left(e^{i\phi}c^\dagger_{{\bf j},\uparrow} c_{{\bf j},\downarrow}+{\rm h.c.}\right)+\\
& \frac{\Omega_{\rm R}}{2}\sum_{{\bf j}}\left(e^{i\pi (j_x+j_y)}c^\dagger_{{\bf j},\uparrow} c_{{\bf j},\downarrow}+{\rm h.c.}\right).
\end{align*}
$c_{{\bf j},s}$ is an annihilation operator (bosonic or fermionic)
on site ${\bf j}$ with spin $s$; the hopping matrix element $t$
can be computed from the band structure of a sinusoidal lattice
(for a $U_s=5\Er$ scalar lattice $t\approx0.07\Er$); $\Omega_{\rm
rf}=g\mu_B B_{\rm rf}$; and $\Omega_{\rm R}=U_v/2$. Since we focus
on very small $\Omega_R\simeq t$, the detuning from atomic
resonance can be quite large.  For $^{40}$K, with
$\Delta_{FS}/2\pi=1.7\THz$, the detuning is
$(\omega_0-\omega)/2\pi\approx50\Delta_{FS} = 86\THz$, yielding a
laser wavelength $\approx980\nm$, far detuned from the $770.1\nm$
(D1) and $766.7\nm$ (D2) transitions.

To determine the single-particle spectrum, we define spinor field
operators
$c_{\mathbf{j}}\equiv(c_{\mathbf{j},\uparrow},c_{\mathbf{j},\downarrow}
)^T$ and three component vectors
$\mathbf{S}_\mathbf{j}=(S^x_\mathbf{j},S^y_\mathbf{j},S^z_\mathbf{j})^T
=c_{\mathbf{j}}^{\dag} \mathbf{\bf\check\sigma} c_{\mathbf{j}}/2$,
where $\mathbf{\check\sigma}$ is the vector of Pauli matrices.
Owing to the staggered Zeeman field, we introduce sublattices
$A_+$ and $A_-$, where
$A_\pm=\left\{(j_x,j_y)\big|(-1)^{j_x+j_y}=\pm1\right\}$ and we
define $a_{\mathbf{j}, \pm}\equiv c_{\mathbf{j}}$ for  $\mathbf{j}
\in A_\pm$. In addition, we introduce vectors
$\mathbf{B}_\pm=(\Omega_{\rm rf}\cos(\phi)\pm\Omega_{\rm
R},-\Omega_{\rm rf}\sin(\phi),0)^T$ describing Zeeman fields on
the $A_\pm$ sublattices. In this notation, the bare Hamiltonian is
\begin{equation*} {H}_0=
-t\sum_{\langle\mathbf{i},\mathbf{j}\rangle}(a^{\dag}_{\mathbf{i,+}}a_{\mathbf{j,-}}+{\rm h.c.})+\sum_{\mathbf{j}\in
A_+}\mathbf{S}^T_\mathbf{j}\cdot \mathbf{B}_+\sum_{\mathbf{j}\in
A_-}\mathbf{S}^T_\mathbf{j}\cdot \mathbf{B}_-.
\end{equation*}
In terms of momentum field operators $\Psi^{\dag}_{\mathbf{k}}=({a}^{\dag}_{\mathbf{k},+,\uparrow},{a}^{\dag}_{\mathbf{k},+,\downarrow},
{a}^{\dag}_{\mathbf{k},-,\uparrow},{a}^{\dag}_{\mathbf{k},-,\downarrow})$, the Hamiltonian becomes
${H}_0=\sum_{\mathbf{k}}\Psi^{\dag}_{\mathbf{k}}\mathbf{H}_{0\;
\mathbf{k}}\Psi_{\mathbf{k}}$ where
\begin{equation*}
\mathbf{H}_{0\; \mathbf{k}}=\left[
\begin{array}{cc} \mathbf{\check\sigma} \cdot \mathbf{B}_+/2 & -t\gamma_\mathbf{k}\mathbf{I}\\
-t\gamma_\mathbf{k}\mathbf{I} & \mathbf{\check\sigma} \cdot
\mathbf{B}_-/2 \nonumber
\end{array} \right].
\end{equation*}
Here, $\gamma_\mathbf{k}=4\cos(k_+/2)\cos(k_-/2)$, with
$k_\pm=\pi(k_x\pm k_y)/\kr$, $\mathbf{k}=(k_+,k_-)$ and $\bf I$ is
the 2$\times$2 identity matrix. The summation goes over the entire
Brillouin zone $-\pi<k_+,k_-\leq\pi$, i.e.,
$\gamma_\mathbf{k}\geq0$. The four eigenvalues of $\mathbf{H}_{0\;
\mathbf{k}}$ are
$\varepsilon_\mathbf{k}^2=(t\gamma_\mathbf{k})^2+(\Omega_{\rm
rf}/2)^2+(\Omega_{\rm R}/2)^2 \pm  \Omega_{\rm
rf}\sqrt{(t\gamma_k)^2+(\Omega_{\rm R}\cos(\phi)/2)^2}$; together
these eigenvalues constitute four bands [Fig.~\ref{geometry}(c)]:
the lowest band has a minimum at $\mathbf{k}=0$, whereas the third
band can be shaped as a squarish, deformed Mexican hat.  The
second band may either exhibit the same trivial behavior as the
first band, with a global minimum at $\mathbf{k}=0$ or have lines
of degenerate minima along a square contour at the edge of the
square Brillouin Zone.  The fourth band always has lines of minima
at the zone boundary.
\begin{figure*}[tb]
\begin{center}
\includegraphics[width=7.1in]{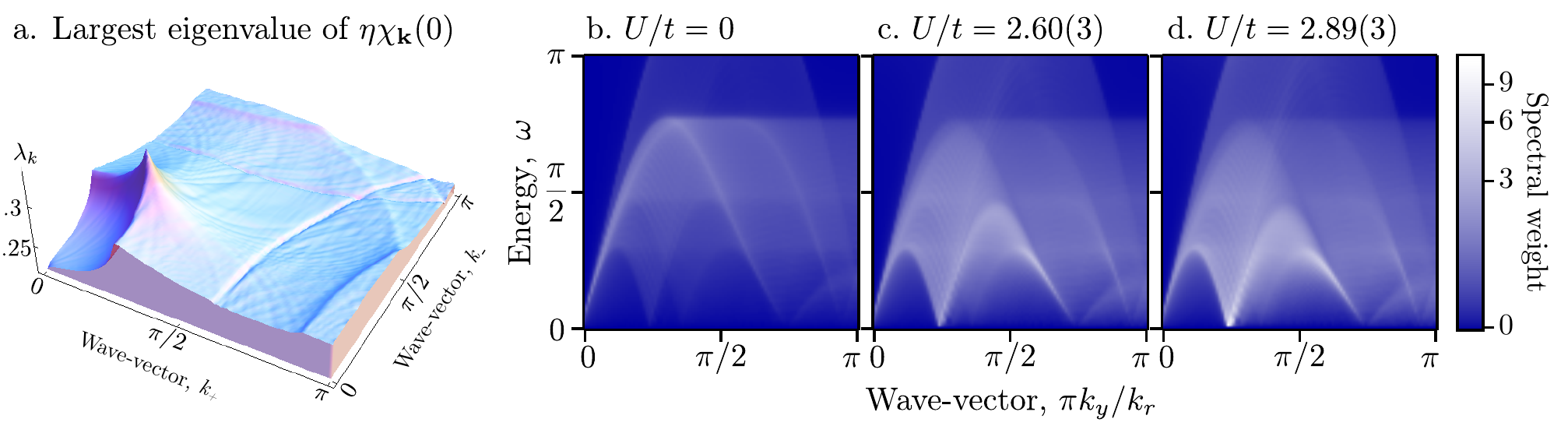}
\end{center}
\caption{\label{Susceptibilities}(a) Largest eigenvalue of the
static susceptibility $\eta\chi_{\mathbf{k}}(0)$ as a function of
$k_+$ and $k_-$.  The peaks indicate the ${\bf Q}$ vectors for
SCDW instabilities, and the largest peak marks the location of the
most prominent nesting vector. Imaginary part of the trace of the
susceptibility ${\rm Tr}[{\rm
Im}\mathbf{\chi}_\mathbf{k}(\omega)]$ in logarithmic scale in the
$k_y$-$\omega$ plane, with $k_x=0$, i.e., $k_+$=$-k_-$. (b)
Without interactions a linearly dispersing sound mode is observed
for small $k_y$. (c) For $U/t = 2.60(3)$, spectral weight builds
up for a second linear-dispersing mode, which starts from
$\omega=0$ at $\pi k_y/\kr=\pi/4$. (d) For $U/t = 2.89(3)$, the sharp
increase of the intensity at $\omega=0$ with $\pi k_y/\kr=\pi/4$ signals
the onset of SCDW instability.}
\end{figure*}

{\it Spin-charge-density-wave} While the
single-particle spectrum is valid for fermions and bosons, we now
focus on spin-$1/2$ fermions with a Hamiltonian
\begin{equation}
H = H_0 + H_{\rm int}; \quad
H_{\rm
int}=U\sum_{\mathbf{j}}c^{\dag}_{\mathbf{j},\uparrow}c^{\dag}_{\mathbf{j},\downarrow}c_{\mathbf{j},\downarrow}c_{\mathbf{j},\uparrow}.
\label{Ha}
\end{equation}
The interaction strength $U$ is proportional to the $s$-wave
scattering length, and the Fermi energy is chosen to be in the
third band.  The resulting squarish Fermi surface is depicted by
the white contours in Fig.~\ref{geometry}(c) and is shaped like
two concentric squircles. Nesting and local fermion-fermion
interactions lead to spin- and charge- ordered phases in this
system.  We anticipate a second order phase transition when the
coefficient of the second order term in the Landau free energy
vanishes.  Formally, we use the Hubbard-Stratonovich
transformation to treat the interactions within a saddle point
approximation (analogous to the time-dependent Hartree-Fock
approximation).

Coexisting spin- and charge-density waves (SDW and CDW) have been
studied in quasi-2D organics (a few coupled chains) using a Monte
Carlo aproach~\cite{Campbell}. Conventional approaches to study
SDW instabilities in the 2D Hubbard model neglect the contribution
of charge density fluctuations~\cite{Fradkin,Negele}, which are
important here.  In the following, we develop a generalized
solution of 2D tight-binding models with local interactions and
obtain a theory of SCDW instabilities.

In the coherent states formalism, the grand-canonical partition
function is $Z=\int \mathrm{d}[c^{\dag}]\mathrm{d}[c]e^{-{\cal
S}[c^{\dag},c]/\hbar}$, where ${\cal S}=\int_0^{\hbar\beta}
\mathrm{d} \tau \left[\sum_{\mathbf{j}} c^{\dag}_{\mathbf{j}}
\left(\hbar \partial_\tau -\mu\right)c_{\mathbf{j}}+H_0+H_{\rm
int}\right] $ is the Euclidean action and $\beta=1/k_BT$. We
express the interaction term in a ${\rm SO}(3,1)$ invariant form
$c^{\dag}_{\mathbf{j},\uparrow}c^{\dag}_{\mathbf{j},\downarrow}c_{\mathbf{j},\downarrow}c_{\mathbf{j},\uparrow}=
(1/8)n_\mathbf{j}^2-(1/2){\mathbf{S}_\mathbf{j}}^T\cdot\mathbf{S}_\mathbf{j}$,
with $n_\mathbf{j}=c_{\mathbf{j}}^{\dag}c_{\mathbf{j}}$. The ${\rm
SO}(3)$ invariance is required by rotational symmetry; the fact
that the spin and charge terms have different signs reflects the
Pauli principle, which requires a vanishing self-energy for a
polarized state.  The Hubbard-Stratonovich transformation renders
the action quadratic in the fermion operators by introducing
auxiliary bosonic fields, $\rho_\pm$ and $\mathbf{M}_\pm$, which
couple to charge and spin density, respectively.  For repulsive
interactions the charge density term leads to a divergent
integral.  We resolve this problem by integrating along a contour
parallel to the imaginary axis for $\rho_\pm$. Next, we introduce
a source field $\mathbf{J}$ that couples to the charge and spin
densities at each sublattice, and an eight-component vector
$\mathbf{M}_{\mathbf{k},n}=(\rho_{\mathbf{k},+,n},
\mathbf{M}_{\mathbf{k},+,n}, \rho_{\mathbf{k},-,n},
\mathbf{M}_{\mathbf{k},-,n})^T$ expressed in terms of momentum
${\bf k}$ and Matsubara frequency $\omega_n=\pi(2n+1)/\hbar\beta$.
After integrating out the fermionic fields, we obtain a
path-integral over the auxiliary bosonic field
$\mathbf{M}_{\mathbf{k},n}$, which we evaluate in the saddle-point
approximation.  Notice that the saddle-point $\langle
\mathbf{M}_{\mathbf{k},n}\rangle_{J}$ depends on the source field
$\mathbf{J}$. We find that $\ud\langle \mathbf{M}\rangle_{J}=\hbar
U\mathbf{\chi}_{J}^{\rm RPA}\ud \mathbf{J}$, where the generalized
RPA susceptibility $(\mathbf{\chi}_{J}^{\rm
RPA})^{-1}=\mathbf{\chi}_{J}^{-1}-\eta U$ is an  $8\times8$
matrix. The matrix $\eta={\rm Diag}(-1,1,1,1,-1,1,1,1)$ is a
metric signature corresponding to the ${\rm SO}(3,1)\times {\rm
SO}(3,1)$ group and $\mathbf{\chi}_{J}$ is the bare susceptibility
for the renormalized Hamiltonian theory.  By neglecting second order fluctuations 
of the Hubbard-Stratonovich fields, we obtain $\ud
\ln(Z[\mathbf{J}])=U^{-1}\langle\mathbf{M}\rangle^{\dag}_{J} \cdot
\ud\mathbf{J}$.  Neglecting now third- and higher-order terms in
$\mathbf{J}$ we find $Z[\mathbf{J}]=Z[0]\exp(-{\cal S_{\rm
eff}[\mathbf{J}]} /\hbar)$, where $ {\cal S}_{\rm eff}= (-\hbar/
U) \langle\mathbf{M}\rangle^{\dag}_{0} \cdot \mathbf{J}- (\hbar^2
/ 2) \mathbf{J}^{\dag} \cdot
(\mathbf{I}-U\mathbf{\chi}^{}_{0}\eta)^{-1}\mathbf{\chi}^{}_{0}
\cdot\mathbf{J} $. The free energy may be determined by performing
a Legendre transformation $\beta F[\langle\mathbf{M}\rangle_{J}]=
U^{-1}\langle\mathbf{M}\rangle^{\dag}_{J} \cdot
\mathbf{J}-\ln(Z[\mathbf{J}])$ (see Ref.~\cite{Negele}) which, up
to quadratic order in the deviation
$\Delta\langle\mathbf{M}\rangle_{J}\equiv\langle\mathbf{M}\rangle_{J}-\langle\mathbf{M}\rangle_{0}$
and without an additive constant reads
\begin{equation}\label{quadraticfreeenergy}
\beta F[\langle\mathbf{M}\rangle_{J}]=\frac{1}{2\hbar U^2}\left(
\Delta\langle\mathbf{M}\rangle_{J}\right)^{\dag} \cdot
(\mathbf{\chi}_{0}^{\rm RPA})^{-1}
\cdot\Delta\langle\mathbf{M}\rangle_{J}.
\end{equation}
The susceptibility $\mathbf{\chi}_{0}^{\rm RPA}$ is evaluated in
the absence of the source field, $\mathbf{J}=0$.  For homogeneous
phases, the susceptibility and the Hamiltonian become diagonal in
momentum and frequency space.  Thus, the Hamiltonian in the
saddle-point approximation becomes
\begin{equation}\label{meanfselfenerg}
\mathbf{H}_{\mathbf{k}}=\mathbf{H}_{0\; \mathbf{k}}-\frac{ U
}{4N}\sum_{\mathbf{q},r,r',\alpha}\mathbf{P}^{ r}\eta^{r,r'}{\rm
Tr}[\mathbf{P}^{ r'}\mathbf{U}_\mathbf{q} \mathbf{I}^{(\alpha)}
\mathbf{U}_\mathbf{q}^{\dag}]n_{\rm F}(\tilde{\varepsilon}_\mathbf{q}^{(\alpha)}).
\end{equation}
$\mathbf{P}^r=[{\rm Diag}(\mathbf{I},0)$, ${\rm
Diag}(\sigma^x,0)$, ${\rm Diag}(\sigma^y,0)$, ${\rm
Diag}(\sigma^z,0)$, ${\rm Diag}(0,\mathbf{I})$, ${\rm
Diag}(0,\sigma^x)$, ${\rm Diag}(0,\sigma^y)$, ${\rm
Diag}(0,\sigma^z)]^T$ are constant $4\times4$ matrices; $N$ is the
number of sites in a sublattice; $n_{\rm F}(z)=(e^{\beta z
}+1)^{-1}$ is the Fermi distribution function; energy is measured
with respect to the chemical potential
$\tilde{\varepsilon}_\mathbf{k}^{(\alpha)}=\varepsilon_\mathbf{k}^{(\alpha)}-\mu$,
and $\mathbf{U}_\mathbf{q}$ is a unitary matrix which diagonalizes
\begin{equation}
\mathbf{H}_\mathbf{k}=\sum_{\alpha}\mathbf{U}_\mathbf{k}
\mathbf{I}^{(\alpha)}
\mathbf{U}_\mathbf{k}^{\dag}\varepsilon_\mathbf{k}^{(\alpha)},
\label{diagH}
\end{equation}
with $\mathbf{I}^{(\alpha)}\equiv {\rm
Diag}(\delta_{\alpha,1},\delta_{\alpha,2},\delta_{\alpha,3},\delta_{\alpha,4})$.
Solving Eqs.\ (\ref{meanfselfenerg}) and (\ref{diagH}) self consistently, we determine
the RPA susceptibility
$(\mathbf{\chi}_{0}^{\rm RPA})^{-1}=\mathbf{\chi}_{0}^{-1}-\eta
U$, where the $k-$dependent susceptibility at zero source $\mathbf{\chi}_{0}$ is
\begin{eqnarray}\label{fullsusceptibility}
\chi^{ r, r'}_{\mathbf{k}}(i\hbar\Omega_n)&=\frac{1}{
N}\sum_{\bf{p},\alpha,\beta
}\frac{n_F(\tilde{\varepsilon}_{\mathbf{p}+\mathbf{k}}^{(\alpha)})-n_F(\tilde{\varepsilon}_{\mathbf{p}}^{(\beta)})}
{\tilde{\varepsilon}_{\mathbf{p}+\mathbf{k}}^{(\alpha)}-\tilde{\varepsilon}_{\mathbf{p}}^{(\beta)}-i\hbar\Omega_n}
T^{r,r';\alpha,\beta}_{\mathbf{p}+\mathbf{k},\mathbf{p}}, \\
\nonumber
T^{r,r';\alpha,\beta}_{\mathbf{p}+\mathbf{k},\mathbf{p}}&
\equiv\frac{1}{2}{\rm
Tr}[\mathbf{P}^{r}\mathbf{U}_{\mathbf{p}+\mathbf{k}}
\mathbf{I}^{(\alpha)}
\mathbf{U}_{\mathbf{p}+\mathbf{k}}^{\dag}\mathbf{P}^{r'}
\mathbf{U}_\mathbf{p}
\mathbf{I}^{(\beta)}\mathbf{U}_\mathbf{p}^{\dag}].
\end{eqnarray}
Here $\Omega_n=2\pi n/\hbar\beta$ is the bosonic Matsubara frequency.

{\it Static susceptibility}
The expression in Eq.\;(\ref{fullsusceptibility})
can be evaluated numerically with arbitrary precision. 
Though our approach is applicable for any temperature
regime, we restrict ourselves to temperatures close to zero.
First, we consider the
static susceptibility $\chi_{\mathbf{k}}(0)$ and
look for possible instabilities.  The instability condition
for repulsive interactions requires $U>U_c$, where $U_c$ is determined by $\det(\mathbf{\chi}^{-1}_{0}-U_c \eta)=0$.
Since we avoid the van Hove singularity, the
susceptibility is finite and the critical value $U_c$ is nonzero.
It is related to the largest eigenvalue $\lambda_{\mathbf{Q}}$ of
the matrix $\eta\chi_{\mathbf{k}}(0)$ by
$U_c^{-1}=\max_{\mathbf{k}}\lambda_{\mathbf{k}}$. Thus, the
instability condition becomes $\lambda_{\mathbf{Q}} U >1$,
analogous to the Stoner criterium.
Fig.~\ref{Susceptibilities}(a) shows $\lambda_{\mathbf{Q}}$ for
the Fermi surface in Fig.~\ref{geometry} ($\phi=\pi/4$,
$\Omega_{\rm R}=2t$, $\Omega_{\rm rf}=4t$, $\mu=t$, and $k_{\rm B}
T=10^{-3} t$). We calculated $\chi_{\mathbf{k}}(0)$ numerically on
each point of a mesh with $240\times240$ points. The peak with
$\lambda_{\mathbf{Q}}=0.345(3)$, corresponding to the critical
value of interactions $U_c/t=2.89(3)$, is located at
$\mathbf{Q}=(\pi/4,\pi/4)$, where we expect an imperfect nesting
between inner and outer lines of the Fermi surface
[Fig.~\ref{geometry}(c)], with
$\varepsilon_{\mathbf{p}+\mathbf{Q}}^{(3)}\approx\varepsilon_{\mathbf{p}}^{(3)}$.
For these system parameters, the eigenvector $V_{\mathbf{Q}}$
corresponding to this eigenvalue has an anti-ferrimagnetic
character and is a mixture of both SDW and CDW, hence a SCDW. The
details of the mixture are not universal.

The period of the SCDW $2\pi/|{\bf Q}|$ is in general {\it
incommensurate} with the lattice period and is {\it freely
tunable} by changing the vectors $\mathbf{B}_\pm$, and the
chemical potential $\mu$. Nesting can also occur for other
momenta, which results into smaller peaks forming the pattern
shown in Fig.~\ref{Susceptibilities}(a).

Had we neglected the coupling with charge and considered only the
spin susceptibility, we would find at the same value of ${\bf Q}$
a much lower value for the critical interaction strength:
$U_c/t=2.13(3)$ compared with $U_c/t=2.89(3)$ in the full
calculation. In addition, when considering only charge
excitations, no CDW instability occurs for repulsive interactions
$U>0$. Thus, the coupling of charge and spin excitations, as
developed here, is essential to the realization of a phenomenon
which otherwise would only occur for attractive interactions
$U<0$.

{\it Collective excitations}
Equation~(\ref{fullsusceptibility}) allows
us to study the collective excitation spectra by analytically
continuing $i\Omega_n\rightarrow \omega+i\kappa$ and looking at
the imaginary part of the trace of the RPA susceptibility ${\rm
Tr}[{\rm Im}\mathbf{\chi}^{\rm RPA}_\mathbf{k}(\omega)]$. For
$\kappa=10^{-2}$ we find a linear dispersion spectrum in
Fig.~\ref{Susceptibilities}(b) in the long wavelength region (the Landau
zero sound), which could have been anticipated,
since we are considering a compressible zero-temperature Fermi
liquid.  At the interaction value $U_c/t=2.89(3)$ a soft linearly
dispersing mode starting from $\mathbf{k}=\mathbf{Q}$ appears,
signaling the onset of instability
[Fig.~\ref{Susceptibilities}(d)]. In $^{40}$K, the collective
excitation spectrum can be experimentally studied with an atomic
analog of angle resolved photoemission spectroscopy~\cite{Stewart2008} or
by measuring the dynamic structure factor with energy and momentum
sensitive Bragg spectroscopy~\cite{Vogels2002}.

{\it Conclusions}
We showed how to construct a system with a
unique Fermi surface consisting of concentric squircles. The
system has peculiar collective excitations, which we analyze in
the RPA including both charge- and spin- density excitations.  Our
studies predict an anti-ferrimagnetic-like instability combining
both CDW and SDW with a tunable incommensurate wave-vector,
determined by the nesting properties of the Fermi surface, for
sufficiently strong interactions. Moreover, we find that the usual
approach -- neglecting the coupling with density fluctuations --
significantly underestimates the critical value of the interaction
strength.

{\it Acknowledgments}
We acknowledge the hospitality of the KITP
Santa Barbara, supported by the National Science Foundation under
Grant No. PHY05-51164, where this work was initiated. I.B.S.
acknowledges the financial support of the ARO with funds from the
DARPA OLE Program, and the NSF through the PFC at JQI. C.M.S. was
partially supported by Netherlands Organization for Scientific
Research NWO.  We are indebted to D. Baeriswyl, G. Baym, A. Hemmerich, 
and A. Lazarides for fruitful discussions.

\end{document}